\documentclass[conference]{IEEEtran}
\IEEEoverridecommandlockouts
\usepackage{cite}
\usepackage{amsmath,amssymb,amsfonts}
\usepackage{algorithmic}
\usepackage{graphicx}
 \usepackage{textcomp}
\usepackage{xcolor}
\usepackage{subcaption}
\def\BibTeX{{\rm B\kern-.05em{\sc i\kern-.025em b}\kern-.08em
 T\kern-.1667em\lower.7ex\hbox{E}\kern-.125emX}}

  \makeatletter
\newcommand*{\rom}[1]{\expandafter\@slowromancap\romannumeral #1@}
\makeatother

\usepackage{xcolor} 

     \usepackage[left= 2.45 cm, right= 2.45 cm, top=1.778 cm, bottom = 2.65 cm]{geometry}

\begin{document}
 \bstctlcite{IEEEexample:BSTcontrol}
 
\title{Adversarial Attacks in AI-Driven RAN Slicing: SLA Violations and Recovery}

\author{\IEEEauthorblockN{Deemah H. Tashman and
Soumaya Cherkaoui }
\IEEEauthorblockA{ \small Department of Computer  and Software Engineering, Polytechnique Montreal, Montreal, Canada\\ 
Email:    \{deemah.tashman, soumaya.cherkaoui\}@polymtl.ca }}

\maketitle
\begin{abstract}
Next-generation (NextG) cellular networks are designed to support emerging applications with diverse data rate and latency requirements, such as immersive multimedia services and large-scale Internet of Things deployments. A key enabling mechanism is radio access network (RAN) slicing, which dynamically partitions radio resources into virtual resource blocks to efficiently serve heterogeneous traffic classes, including  enhanced mobile broadband (eMBB), massive machine-type communications (mMTC), and ultra-reliable low-latency communications (URLLC). In this paper, we study the impact of adversarial attacks on AI-driven RAN slicing decisions, where a budget-constrained adversary selectively jams slice transmissions to bias deep reinforcement learning (DRL)–based resource allocation, and quantify the resulting service level agreement (SLA) violations and post-attack recovery behavior. Our results indicate that budget-constrained adversarial jamming can induce severe and slice-dependent steady-state SLA violations. Moreover, the DRL agent’s reward converges toward the clean baseline only after a non-negligible recovery period.
\end{abstract}
\begin{IEEEkeywords}
Adversarial attacks, deep reinforcement learning, radio access network slicing,   service level agreement.
\end{IEEEkeywords}
\section{Introduction}

\par\IEEEPARstart{T}{o} support applications such as enhanced mobile broadband (eMBB), massive machine-type communications (mMTC), and ultra-reliable low-latency communications (URLLC), next-generation (NextG) networks introduce major architectural enhancements through radio access network (RAN) slicing mechanisms \cite{10756600,10433640,11178232,10921752,10078092,9999295,9318243}. In this paradigm, physical radio resources are virtualized and partitioned into resource blocks (RBs) that can be dynamically allocated across slices, enabling  slice-aware resource management via intelligent RAN controllers \cite{11124199,sarker2026network,9729992}. 

Service level agreements (SLAs) are crucial to RAN slicing since they provide the anticipated performance guarantees between network operators and tenants in terms of key performance indicators (KPIs), such as latency and throughput \cite{yungaicela2025rslaq}. Adhering to the SLA is crucial not just for maintaining satisfactory service, but also for mitigating penalties, safeguarding reputation, and minimizing regulatory concerns, particularly for services that are sensitive to latency and reliability. NextG   architectures employ intelligent RAN controllers   to translate high-level SLA requirements into granular resource allocation decisions \cite{yungaicela2025rslaq}. Moreover, to enable intelligent resource allocation, artificial intelligence (AI) has been increasingly integrated into critical RAN functions, facilitating adaptive decision-making through learning from network measurements and dynamically optimizing slice-level RB assignments \cite{9904606,11370849}.

Despite the benefits of AI-driven control in  RANs, the reliability of learning-based decision mechanisms is increasingly challenged by adversarial threats \cite{11371394,10623131}.  By deliberately manipulating uplink traffic patterns, network measurements, or feedback signals, an adversary can distort the state information observed by AI-based resource allocation agents \cite{10437043,10437829,11431904}. Such interference can bias the learned policies toward suboptimal or harmful actions, leading to incorrect slice-level resource block assignments, persistent SLA violations, and, in extreme cases, service disruption across multiple network slices.  For instance, reinforcement learning (RL) systems continuously interact with the environment and update their policies based on observed rewards \cite{10646359,10182973}. This adaptive behavior gives rise to   properties that can be exploited by an adversary.  That is, it becomes feasible  to manipulate  the reward signal to significantly degrade the learning and decision-making performance of the   agent. For example, the authors in \cite{lacava2025poison} evaluated data poisoning and backdoor attacks against deep reinforcement learning (DRL)-based xApps in open RAN (O-RAN), showing that poisoned training data can manipulate control decisions and severely degrade network performance. Moreover, \cite{10697477} studied adversarial attacks against DRL-based resource allocation in the Near-Real-Time RAN Intelligent Controller (RIC) for vehicular networks, showing that manipulated observations can mislead AI agents and severely degrade RB allocation, data rates, latency, and URLLC reliability. In \cite{aizikovich2025rogue}, the authors analyzed adversarial attacks against AI-enabled O-RAN control in multi-operator deployments, showing that a malicious cell can spoof KPIs to deceive traffic steering decisions and gain unfair RB allocations. Furthermore, the authors in \cite{10971998} explored O-RAN  adversarial attacks, including policy infiltration, gradient-based attacks, and signal perturbation. Finally, \cite{9984930}  studied adversarial over-the-air attacks against RL-based network slicing in NextG RANs, where an intelligent jammer selectively disrupts RBs to manipulate the RL reward signal and cause long-term performance degradation.

 While existing works emphasize the development of specific attacks or defense mechanisms, they do not provide a systematic analysis of the impact of adversarial manipulation on SLA outcomes. Hence, to the best of our knowledge, this is the first work to  analyze the impact of adversarial manipulation on AI-driven RAN slicing decisions, with a focus on   SLA violations, steady-state degradation, and post-attack recovery behavior. Given this, the main contributions of this work are given as follows:
\begin{itemize}
    \item We propose a slice-level NextG RAN slicing framework that captures RB allocation, slice activity, and SLA constraints for heterogeneous services, including eMBB, URLLC, and mMTC.
    
    \item We investigate an adversarial DRL–based attack that selectively disrupts slice transmissions under a constrained jamming budget, and analyze its impact on the learning dynamics and decision-making behavior of DRL-driven RAN slicing.
    
    \item We provide a systematic slice-level evaluation of adversarial effects, quantifying steady-state SLA violations and performance degradation across different slice configurations.
    
    \item We analyze post-attack recovery behavior after attack removal, highlighting slice-dependent resilience and long-term performance impairment in AI-driven RAN slicing.
\end{itemize}

The remainder of the paper is organized as follows. Section~\rom{2} presents the system model. Section~\rom{3} formulates the optimization problem and describes the proposed DRL-based solution. The adversarial attack model against the victim agent is introduced in Section~\rom{4}. Simulation results and performance evaluation are provided in Section~\rom{5}, while conclusions are drawn in Section~\rom{6}

\section{System Model}
\subsection{Network Architecture and Slicing Model}
\begin{figure} 
  \centering
  \includegraphics[width=1.2\linewidth]{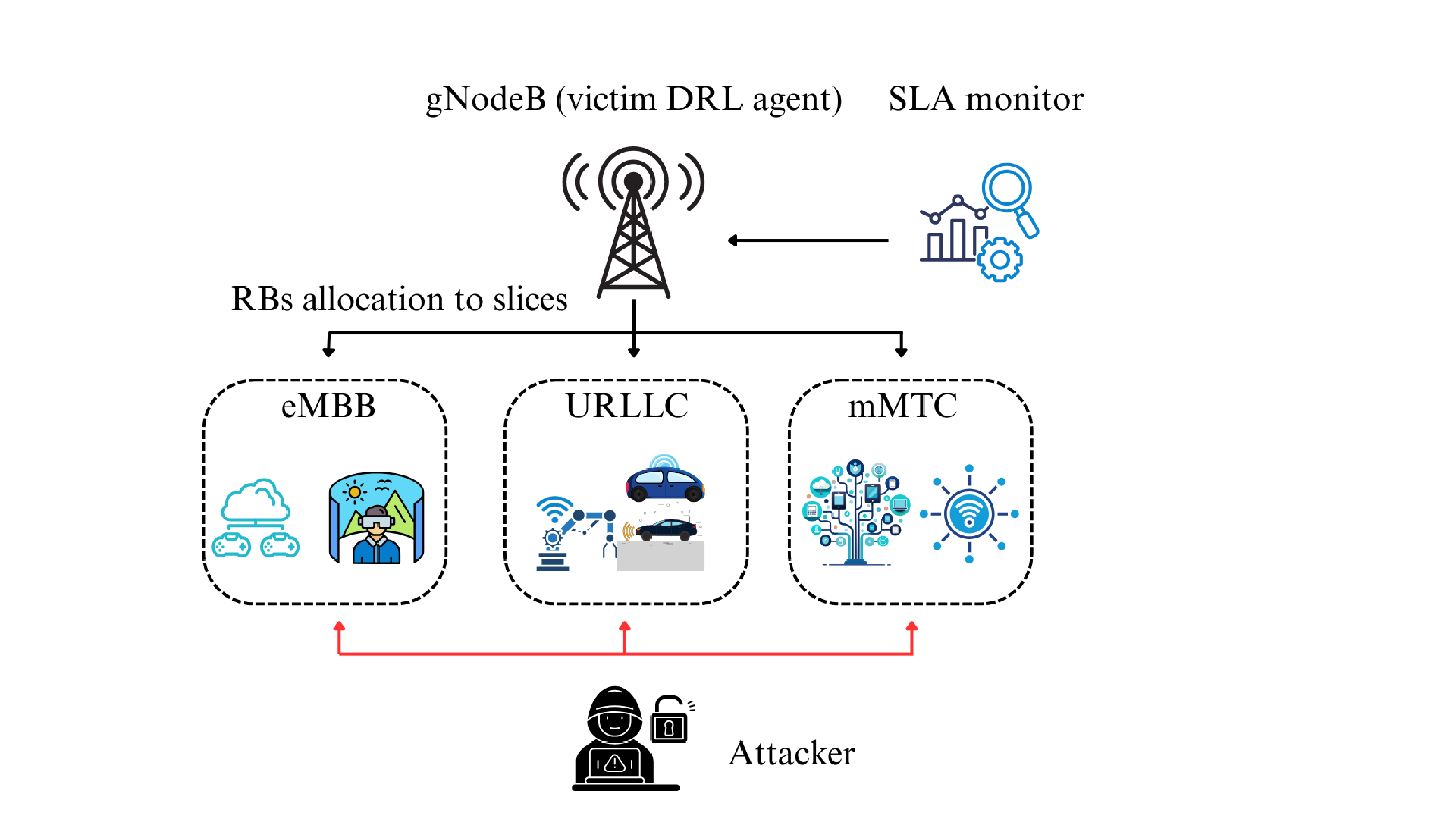}
  \caption{The system model.}
  \label{sys1}
\end{figure}

Fig. \ref{sys1} illustrates a single-cell downlink RAN in which a gNodeB serves multiple network slices, encompassing eMBB, URLLC, and mMTC. We assume that the time is divided into slots, and within each slot, each slice may submit a transmission request based on a stochastic arrival process. When operational, a slice aggregates traffic from multiple user equipment (UEs) with similar quality of equipment (QoE) requirements, including minimum data rate and service reliability, and is assigned to a priority weight that reflects its relative significance. The gNodeB allocates a limited quantity of RBs to a subset of active slices, adhering to an overall RB capacity limitation. The objective is to optimize long-term system utility while adhering to SLAs. 
In addition, the system faces an external adversary that selectively disrupts the transmissions of specific slices within a constrained jamming budget, therefore impairing their service performance.   


Let $\mathcal{K} = \{1,2,3\}$ denote the set of network slices corresponding to eMBB, URLLC, and mMTC, respectively. At each time slot $t$, slice $k \in \mathcal{K}$ generates a traffic request with probability $p_k$. We define a binary activity indicator as
\begin{equation}
a_k(t) =
\begin{cases}
1, & \text{if slice } k \text{ has an active request at time } t, \\
0, & \text{otherwise}.
\end{cases}
\end{equation}
The set of active slices at time $t$ is given by $\mathcal{A}(t) = \{ k \in \mathcal{K} \mid a_k(t) = 1 \}$.

Each active slice is associated with a priority weight $w_k(t)$, drawn from a bounded interval, reflecting its relative importance. Moreover, each slice $k$ requires a fixed number of RBs, denoted by $F_k$, to be served in a given time slot.

\subsection{Downlink Data Rate and QoE Constraints}

For each active slice $k \in \mathcal{A}(t)$, the achieved downlink data rate at time $t$ is approximated as \cite{9984930}
\begin{equation}
D_k(t) = c \, F_k \, \big(1 - \mathrm{BER}_k(t)\big),
\label{eq:datarate}
\end{equation}
where $c$ is a constant determined by the modulation and bandwidth configuration, and $\mathrm{BER}_k(t)$ denotes the bit error rate of slice $k$, modeled as a function of the signal-to-noise ratio at user $k$  under the AWGN channel assumption with the adopted modulation and coding scheme.

Each slice has a minimum rate requirement $R_k^{\min}$. Therefore, a slice satisfies its instantaneous rate constraint if
\begin{equation}
D_k(t) \ge R_k^{\min}.
\end{equation}

\subsection{Resource Allocation Constraints}

The gNodeB has a total of $F$ available RBs per time slot. Let $x_k(t) \in \{0,1\}$ be a binary decision variable indicating whether slice $k$ is served at time $t$. Given this, the resource allocation must satisfy the following constraint \cite{9984930}
\begin{equation}
\sum_{k \in \mathcal{A}(t)} F_k \, x_k(t) \le F.
\label{eq:rb_constraint}
\end{equation}
\noindent This constraint ensures that the total number of RBs allocated to the set of active and scheduled slices at time slot $t$ does not exceed the gNodeB’s available RB capacity.

\subsection{SLA Formulation and Evaluation Metrics}
\subsubsection{Served-Ratio SLA}
The served ratio captures the fraction of slice transmission requests that are successfully served within a finite observation window, and is used as a slice-level SLA metric. We further define a binary success indicator $s_k(t)\in\{0,1\}$, where $s_k(t)=1$ if slice $k$ is successfully served at time $t$, i.e., it is scheduled ($x_k(t)=1$), not jammed, and its achieved rate satisfies the minimum rate requirement; otherwise $s_k(t)=0$.
 For slice $k$, the served ratio over the window ending at time $t$ is defined as
\begin{equation}
\rho_k(t) =
\frac{
\sum_{\tau=t-W+1}^{t} \mathbb{I}\{s_k(\tau)=1 \land a_k(\tau)=1\}
}{
\sum_{\tau=t-W+1}^{t} \mathbb{I}\{a_k(\tau)=1\}
},
\label{eq:served_ratio}
\end{equation}
where $\mathbb{I}\{\cdot\}$ denotes the indicator function and $W$ denotes the length of the sliding observation window (in time slots). We also assume that each slice has a minimum served-ratio requirement of  $\rho_k^{\min}$.

\subsubsection{Average Rate SLA}
The average achieved data rate of slice $k$ over the same window is given by
\begin{equation}
\bar{D}_k(t) =
\frac{1}{|\mathcal{T}_k|}
\sum_{\tau \in \mathcal{T}_k} D_k(\tau),
\label{eq:avg_rate}
\end{equation}
where $\mathcal{T}_k$ denotes the set of time slots within the window in which slice $k$ is active.


\subsubsection{SLA Satisfaction Indicator}
A slice is considered to satisfy its SLA if both of the following conditions hold within a sliding observation window:
\begin{itemize}
\item its average achieved data rate exceeds a predefined minimum rate threshold, and
\item the fraction of time slots in which its requests are successfully served exceeds a target served-ratio.
\end{itemize}
Hence, the SLA indicator is given as
\begin{equation}
\mathrm{SLA}_k(t) =
\mathbb{I}\!\left[
\bar{D}_k(t) \ge R_k^{\min}
\;\land\;
\rho_k(t) \ge \rho_k^{\min}
\right].
\label{eq:sla_flag}
\end{equation}




\section{Problem Formulation and Proposed DRL-based Solution}

\subsection{Problem Formulation}
The gNodeB aims to dynamically allocate radio resources among active network slices  to maximize long-term system utility while maintaining slice-level service reliability. Therefore, we formulate a slice-level resource allocation problem that maximizes the weighted sum of successfully served slice requests, while respecting the SLA constraints given in (\ref{eq:sla_flag}). The optimization is subject to per-slot radio RB capacity constraints and binary scheduling decisions for each slice. Given this, the slice-level resource allocation problem can be formulated as
\begin{equation}
\max_{\{x_k(t)\}} \;
\sum_{t}\sum_{k\in\mathcal{A}(t)} w_k(t)\, s_k(t),
\label{eq:opt_problem}
\end{equation}
subject to
\begin{align}
& \sum_{k\in\mathcal{A}(t)} F_k\,x_k(t) \le F, \qquad \forall t, \label{eq:opt_rb}\\
& x_k(t)\in\{0,1\}, \qquad \forall k,\forall t, \label{eq:opt_binary}\\
& \rho_k(t) \ge \rho_k^{\min}, \qquad \forall k,\ \forall t \ge W, \label{eq:opt_served_ratio}\\
& \bar{D}_k(t) \ge R_k^{\min}, \qquad \forall k,\ \forall t \ge W. \label{eq:opt_avg_rate}
\end{align}

\subsection{Deep Reinforcement Learning-based Solution}

To tackle the optimization problem in (\ref{eq:opt_problem}), we employ a   DRL framework  at the gNodeB. Tabular Q-learning has become obsolete due to the expanding state space resulting from dynamic traffic arrivals and varying slice requirements. Hence, we employ a double deep Q-network (DDQN) architecture to approximate the optimal action-value function.

At each time slot $t$, the gNodeB observes the system state $s_t \in \mathcal{S}$, selects an action $a_t \in \mathcal{A}$, receives a scalar reward $r_t$, and transitions to the next state $s_{t+1}$. The state vector captures slice-level information and is defined as
\begin{equation} \label{state-victim}
s_t = \big[ F(t), \; a_1(t), w_1(t), \ldots, a_K(t), w_K(t) \big],
\end{equation}
where $F(t)$ denotes the number of available RBs, $a_i(t) \in \{0,1\}$ indicates whether slice $i$ has an active request, and $w_i(t)$ is the priority weight associated with slice $i$, for $i\in\mathcal{K}$.

The action space consists of all possible subsets of slices that can be served in a given time slot and is represented as a binary bitmask over slices,
\begin{equation}
\mathcal{A} = \{0,1\}^\mathcal{K},
\end{equation}
where a value of $1$   indicates that the slice is selected for service. Infeasible actions (violating the RB capacity) result in zero service in that slot.

The reward function is designed to jointly capture throughput maximization and SLA satisfaction. Specifically, the instantaneous reward at time $t$ is defined as
\begin{equation}
r_t = \sum_{k\in\mathcal{A}(t)} w_k(t)\, s_k(t)
\;-\;
\lambda \sum_{k\in\mathcal{K}} \big(1-\mathrm{SLA}_k(t)\big),
\label{eq:reward}
\end{equation}
\noindent where $\lambda$ is the SLA penalty coefficient, and $\mathrm{SLA}_k(t)\in\{0,1\}$ indicates whether slice $k$ satisfies its SLA over the sliding window. The proposed reward formulation incentivizes the gNodeB to balance short-term utility maximization with long-term SLA compliance across heterogeneous network slices.

The DDQN framework maintains two neural networks: an online network with parameters $\theta$ and a target network with parameters $\theta^-$. The action-value function is updated according to 
\begin{IEEEeqnarray}{rCl}
Q_{\theta}(s_t,a_t) &\leftarrow& Q_{\theta}(s_t,a_t) + \alpha \Big( r_t  + \gamma \nonumber\\
&& Q_{\theta^-}\big(s_{t+1}, \arg\max_{a'} Q_{\theta}(s_{t+1},a')\big)
- Q_{\theta}(s_t,a_t) \Big), \nonumber \\
\end{IEEEeqnarray}
where $\alpha$ is the learning rate and $\gamma$ is the discount factor. This DDQN formulation mitigates the overestimation bias commonly observed in standard DQN.

\section{Deep Reinforcement Learning--Based Slice-Level Attack}

The adaptive behavior of the DRL agent heightens the risk of various adversarial attacks. In this paper, we investigate the reward manipulation attack on DRL-based RAN slicing and examine its substantial effects on the agent's learning and decision-making efficiency. Furthermore, we demonstrate that once the attack is removed, the DRL agent gradually recovers through continued interaction with the environment \cite{10921906,10794361,10622458,10592377}, leading to the restoration of long-term SLA satisfaction.

\paragraph{Attack Surface}
As demonstrated in (\ref{state-victim}), the state observed by the gNodeB encompasses the available RBs, slice activity indicators, and slice priority weights. The gNodeB internally maintains these quantities, preventing a wireless attacker from directly accessing or manipulating the victim agent's state or learning process. 
The attacker can, however, alter the gNodeB's  reward  by actively jamming specific slice transmissions \cite{9984930}. By disrupting successful transmissions, jamming directly affects SLA satisfaction indicators, resulting in diminished rewards and misleading feedback to the learning agent.  In addition, when a slice transmission is jammed, the associated request fails to satisfy its quality of service (QoS) requirements. This leads to a negative acknowledgment (NACK) to be sent from the UEs.

\paragraph{Limited Jamming Capability}
We assume a practical adversary with a constrained jamming budget, indicating restrictions on either the finite transmission power or the energy availability. Let $B$ be the maximum number of RBs that the adversary is capable to jam within a specified time interval. It is worth mentioning that the  adversary cannot disrupt all slices simultaneously, as each slice employs a predefined quantity of RBs. Hence, the adversary must carefully select which slices to target within this budgetary constraint.

\paragraph{Surrogate Attacker Model}
In an optimal situation, the attacker would precisely identify the slices that the gNodeB will serve and would disrupt them accordingly. This type of knowledge is not available in reality, as the adversary is unaware of the slice weights, scheduling choices, or internal gNodeB states. To mitigate this limitation, we employ a \emph{surrogate learning} methodology as outlined in \cite{9984930}. The adversary develops an independent DDQN-based surrogate agent to replicate the behavior of the victim gNodeB. The surrogate does not replicate the victim's precise strategy; nevertheless, it acquires statistical patterns in slice scheduling and resource availability, thereby facilitating informed attack decisions.

\paragraph{Attacker State, Action, and Reward}
The DRL-based attacker is modeled as follows:
\begin{itemize}
    \item \textbf{State}: Similar to \cite{9984930}, we assume that the attacker observes the available RBs at   gNodeB at each time interval   by passively observing slice-level transmission and scheduling activity on the radio interface.
    \item \textbf{Action}: The attacker selects a binary action over slices, indicating which slices to jam. The selected slices must adhere to the jamming budget constraint, such that the cumulative RB demand of the jammed slices does not exceed $B$. 
    \item \textbf{Reward}: The attacker receives a reward equals the number of victim slice transmissions that fail due to jamming, as inferred from observed NACKs. The adversary does not decode the NACK contents but only detects their presence, which is feasible since NACK signals are shorter and structurally distinguishable from data transmissions.
\end{itemize}

\paragraph{Learning Algorithm}
The attacker uses a DDQN to mitigate overestimation bias and enhance learning stability.   The surrogate attacker is trained against a static victim policy, enabling it to learn jamming techniques without direct access to the gNodeB's internal parameters. This DRL-based surrogate attack enables the adversary to systematically diminish slice-level QoS and induce SLA violations, even within stringent jamming budget limitations.

\section{Simulation results}
In this section, we present the simulation setup for the proposed scenario. Unless otherwise stated, all slice-dependent parameters follow the slice ordering \{eMBB, URLLC, mMTC\}. The parameters  of the simulation are set to the following values: The minimum rate requirement for each slice is set to 80\% of its nominal achievable rate, i.e., $R_k^{\min}=0.8 c F_k$, $F=11$, $B=5$, $c= 12.59 \times 10^6$,  $\lambda=2$, $p_k= [0.6,0.4,0.3]$, $w_k = [0.6, 0.4, 0.3]$, $\rho_k^{\min}= [0.85, 0.95, 0.80]$, $F_k= [5,3,1]$, $W=20$, $\gamma=0.9$, and $\alpha=0.1$.

  

 
 \begin{figure}   
  \centering
  \includegraphics[width=1.0\linewidth]{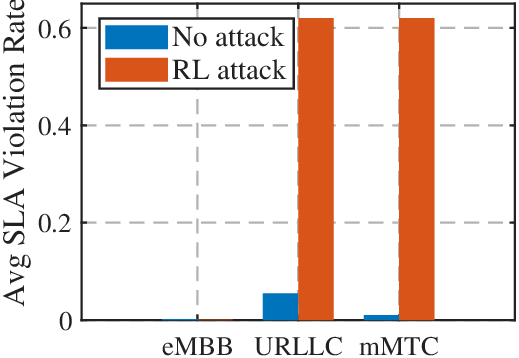}
  \caption{Slice-level average SLA violation rate under no attack and DRL-based adversarial attack}
  \label{fig1}
\end{figure}

Fig. \ref{fig1} illustrates the steady-state slice-level SLA violation rates under both the no-attack scenario and the proposed DRL-based jamming attack. In the absence of attacks, all slices predominantly meet their SLA criteria, with only negligible violations resulting from stochastic traffic arrivals and temporary resource contention. When the NextG network is under attack, significant SLA violations occur, particularly affecting the URLLC and mMTC slices. However, the impact on eMBB remains minimal. Given the limited jamming budget $(B=5)$, and the higher RB demand of eMBB $(F_{\text{eMBB}}=5)$, the learned attack policy avoids eMBB and instead targets URLLC and mMTC, where jamming yields more frequent SLA violations per unit of jamming effort. Additionally, we notice that URLLC and mMTC exhibit nearly identical steady-state SLA violation rates as their combined RB demands remain within the attacker’s jamming budget, allowing both slices to be consistently disrupted and driving their SLA metrics to similar saturation levels.  This behavior underscores the attacker’s capacity to carefully manipulate the victim agent’s SLA-aware reward framework by consistently degrading slices with service guarantees that are more susceptible to missed transmissions and accumulated SLA penalties.  

\begin{figure}   
  \centering
  \includegraphics[width=1.0\linewidth]{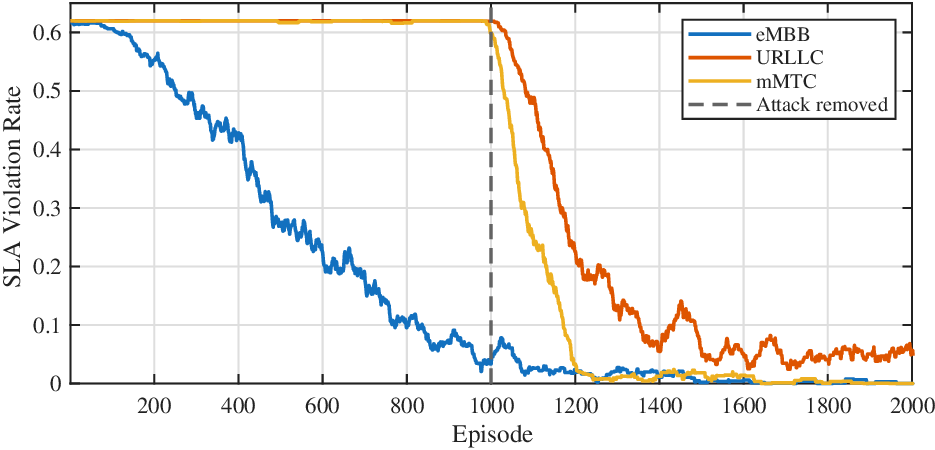}
  \caption{Post-attack SLA recovery behavior per network slice under an RL-based adversarial attack.}
  \label{fig2}
\end{figure}

Fig. \ref{fig2} illustrates the recovery of slice-level SLA violations following the termination of the DRL-based jamming attack. Throughout the attack phase, all slices exhibit persistently elevated SLA violation rates. This indicates that the attacker can prevent the victim from fulfilling long-term QoE needs. Post-attack, the slices exhibit a distinct and visible recovery pattern. The eMBB slice exhibits the most rapid recovery, with its SLA violation rate swiftly declining to zero. On the other hand, URLLC and mMTC, necessitate a longer recovery period due to their elevated SLA thresholds and window-based assessment, which demand successful resource allocation to proceed before SLA violations can be cleared.  The consistent decline in SLA breaches following the removal of the attack indicates that the victim's DRL policy is not permanently harmed by the attack and can readapt through continious learning.

\begin{figure}   
  \centering
  \includegraphics[width=1.0\linewidth]{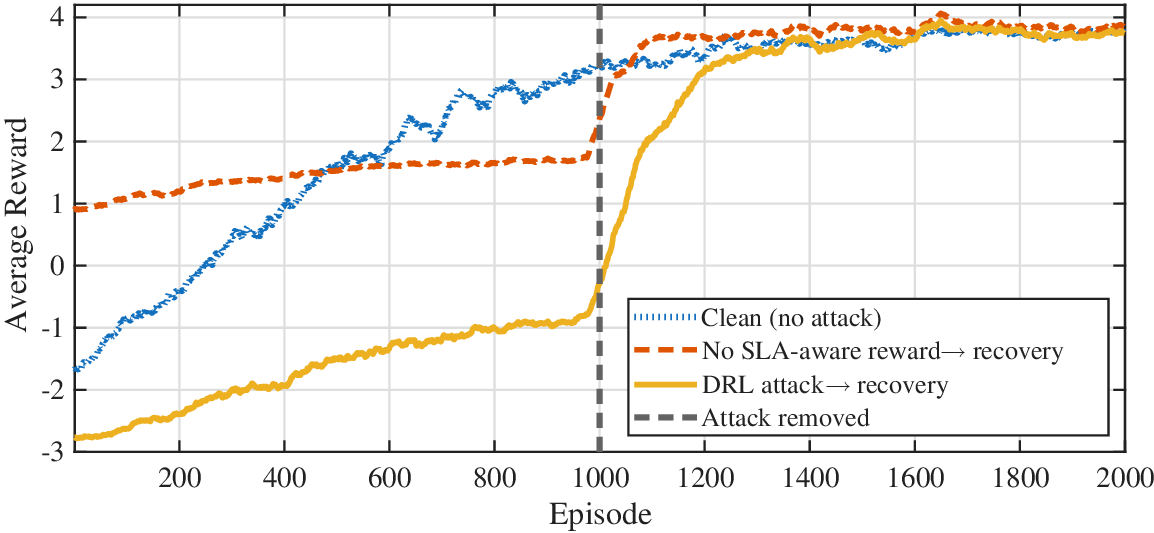}
  \caption{Post-attack recovery of the average reward per slot under an RL-based adversarial attack. }
  \label{fig4}
\end{figure}

Fig. \ref{fig4} illustrates the temporal variation in the victim's average reward per slot during the proposed DRL-based jamming attack and recovery, contrasted with a clean baseline trained without any adversarial interference. The victim's reward is greatly reduced during the attack phase. This results from both the immediate transmission failures due to jamming and the accumulated SLA penalties within the reward function. We also compare our approach with the method proposed in \cite{9984930}, whose objective formulation does not explicitly account for SLA constraints. We conclude that  the benchmark   achieves higher absolute reward values,   primarily due to its SLA-unaware objective, which does not penalize persistent violations of service reliability and continuity. In contrast, the proposed SLA-aware formulation jointly captures throughput and long-term reliability requirements, providing a stricter and more realistic assessment of network performance and revealing degradation that would otherwise remain hidden under throughput-only reward metrics. Subsequent to the termination of the attack, there is a significant change in the victim's policy, which rapidly improves through ongoing engagement with the environment. The restored reward trajectory aligns with the clean baseline, signifying that the RL-based slicing agent may reacquire the ability to allocate resources, even after prolonged exposure to attacks. The temporary gap between the recovery and clean curves immediately following the completion of the attack illustrates the extent of harm that adversarial disruption can inflict; however, the ultimate convergence of the two curves indicates that the attack does not result in permanent damage to the policy.

\section{Conclusions}
In this paper, we investigated adversarial attacks against a DDQN-based RAN slicing framework designed to maximize resource allocation efficiency for heterogeneous services, namely eMBB, URLLC, and mMTC. The adversary was modeled as an intelligent DRL agent that mimics the victim DDQN policy in order to maximize its jamming impact on slice transmissions. By selectively disrupting slice-level communications, the attacker  manipulates the resource allocation process, leading to significant violations of slice-level SLAs. 
Our results demonstrate the substantial effectiveness of the proposed DRL-based attacker in inducing sustained SLA violations by anticipating the victim agent’s allocation decisions. We further show that the eMBB slice recovers more rapidly than URLLC and mMTC after attack removal, as the learned attack policy primarily targets URLLC and mMTC to greedily maximize the number of disrupted slices under a constrained jamming budget, while largely avoiding the higher-cost eMBB slice.  Across all slices, the DRL-based resource allocation agent is ultimately able to reduce SLA violation rates once the attack ceases, indicating a degree of resilience in the learning process. In addition, we observe that adversarial jamming significantly degrades the agent’s average reward due to both transmission failures and accumulated SLA penalties; however, following attack removal, the agent progressively relearns effective allocation strategies and converges back toward the clean, no-attack performance baseline.


\end{document}